\documentclass{an}
\usepackage{times}
\usepackage{graphicx}
\usepackage{subfigure}
\usepackage{epsfig}
\begin{document}
\title{The visibility of low-frequency solar acoustic modes}

\author{A.~M. Broomhall\inst{1}, W.~J. Chaplin\inst{1},
Y. Elsworth\inst{1}, S.T. Fletcher\inst{2}}
\titlerunning{Visibility of p modes}
\authorrunning{A.M. Broomhall et al.}
\institute{School of Physics and Astronomy, University of
Birmingham, Edgbaston, Birmingham B15 2TT \and Faculty of Arts,
Computing, Engineering and Sciences, Sheffield Hallam University,
Sheffield, S1 1WB}


\abstract{We make predictions of the detectability of low-frequency
p modes. Estimates of the powers and damping times of these
low-frequency modes are found by extrapolating the observed powers
and widths of higher-frequency modes with large observed
signal-to-noise ratios. The extrapolations predict that the
low-frequency modes will have small signal-to-noise ratios and
narrow widths in a frequency-power spectrum. Monte Carlo simulations
were then performed where timeseries containing mode signals and
normally distributed Gaussian noise were produced. The mode signals
were simulated to have the powers and damping times predicted by the
extrapolations. Various statistical tests were then performed on the
frequency-amplitude spectra formed from these timeseries to
investigate the fraction of spectra in which the modes could be
detected. The results of these simulations were then compared to the
number of p-modes candidates observed in real Sun-as-a-star data at
low frequencies. The fraction of simulated spectra in which modes
were detected decreases rapidly as the frequency of modes decreases
and so the fraction of simulations in which the low-frequency modes
were detected was very small. However, increasing the
signal-to-noise (S/N) ratio of the low-frequency modes by a factor
of 2 above the extrapolated values led to significantly more
detections. Therefore efforts should continue to further improve the
quality of solar data that is currently available.}

\maketitle

\section{Introduction}\label{section[introduction]}
To date a multitude of solar acoustic (p) modes have been observed
over a wide range of frequencies. However, no independently
confirmed detections of low-degree (low-$l$) p modes, with
frequencies below $\sim970\,\rm\mu$Hz, have been made (e.g.
\cite{Garcia2001}; \cite{Chaplin2002}; \cite{Broomhall2007}). As
low-frequency p modes are expected to have very long lifetimes their
detection would allow their frequencies to be measured to very high
accuracies and precisions. This is crucially important as the
properties of low-$l$ p modes are affected by conditions deep in the
solar interior and so their frequencies act as a probe of these
regions. However, the signal from low-frequency p modes is very
weak. Therefore, different analysis procedures have been developed
in attempts to detect these modes.

Broomhall et al. (2007)\nocite{Broomhall2007} used statistical
techniques to look for coincident prominent features in BiSON and
GOLF frequency-amplitude spectra. They found that many low-frequency
p modes remain undetected despite reducing amplitude detection
thresholds to less than $3$\,mm\,s$^{-1}$\,bin$^{-1/2}$. However,
Broomhall et al.\nocite{Broomhall2007} found the $l=0$, $n=6$ mode
at $\sim973\,\rm\mu Hz$ to be extremely prominent. This mode has
also been detected in other studies such as Garcia et al.
(2001)\nocite{Garcia2001} and Chaplin et al.
(2002)\nocite{Chaplin2002}. The prominence of this mode is
conspicuous because of the lack of evidence for higher-frequency
modes such as the $l=0$, $n=7$ mode at $\sim1118\,\rm\mu Hz$. As yet
this mode has only been observed in Sun-as-a-star data by Garcia et
al. (2001)\nocite{Garcia2001}. These higher-frequency modes should
be easier to observe as theoretically they should exhibit larger
amplitudes in the photosphere, where the observations are made. It
is, therefore, of interest to investigate how prominent the $l=0$,
$n=6$ mode, and the modes surrounding it in frequency, should be.
This will provide an indication as to whether we should be observing
modes that we are not; or whether some effect makes the $l=0$, $n=6$
mode more prominent than the rest.

How easy low-frequency modes are to detect depends upon their power
and width in a frequency-power spectrum. As many low-frequency modes
have not yet been detected their powers cannot be determined from
solar data directly. Here, predictions of the power in very
low-frequency modes have been made by extrapolating the power
observed in well-defined, higher-frequency modes. The height of a
mode in a frequency spectrum also depends upon the width in
frequency over which the power is spread. As the power in the signal
from a mode is damped over time the signal from the mode may be
spread over several bins of a frequency spectrum. The lifetime of a
mode varies with frequency and so it is also possible to infer from
extrapolation the width a mode at a particular frequency is expected
to exhibit from results for well-observed modes. The details of how
the extrapolations were made will be explained in Section
\ref{section[extrapolations]}. To make the extrapolations we have
assumed that the simple functional relationships that turn out to
describe the powers and heights of the well-observed modes are still
valid at low frequencies. The validity of this assumption was tested
using predictions of the powers and widths of low-frequency modes
made by the Cambridge stochastic excitation and damping codes (e.g.
Houdek et al. 1999\nocite{Houdek1999}). Monte Carlo simulations
based on the results of the extrapolations were then performed to
investigate how often simulated modes could be detected (Section
\ref{section[simulations]}). The simulations involved creating
timeseries to mimic Sun-as-a-star data. The simulated data contained
mode signals that have the properties predicted by the
extrapolations. Various statistical tests, which are described in
Section \ref{section[stat tests]}, were performed on the
frequency-amp\-litude spectra that were created from the simulated
timeseries. The statistical tests followed those described in
Broo\-mhall et al. (2007)\nocite{Broomhall2007} and determined how
often the simulated modes could be detected. The results of these
tests are detailed in Section \ref{section[results]}. A discussion
of the results is presented in Section \ref{section[discussion]}. In
this discussion we pay particular attention to the unusual $l=0$,
$n=6$ radial mode.

\section{Predicting the Widths and Powers of
modes}\label{section[extrapolations]} Various properties of modes
can be found by fitting frequen\-cy-power spectra. These properties
include the width of a mode, and therefore its damping time, and the
height of a mode, i.e. its maximum power spectral density. The
product of the height and width is proportional to the total power
in the mode. A BiSON spectrum, consisting of 3071 d of Sun-as-a-star
Doppler velocity observations, was fitted using the methods
described in \cite{Fletcher2007}. The fitting procedure involved
taking the Fourier transform of the timeseries to produce a
frequency-power spectrum and then fitting a Lorentzian-like model to
the various mode peaks in the resulting frequency-power spectrum.
The observations were made between 1996 April 20 and 2004 September
15, an epoch which spans most of solar activity cycle 23. The data
were processed in the manner described by \cite{Appourchaux2000} and
Chaplin et al. (2002)\nocite{Chaplin2002}. The data were the same as
the BiSON data used in Broomhall et al. (2007)\nocite{Broomhall2007}
except that, instead of being rebinned to have a cadence of
$120\,\rm s$, the data were left with the nominal $40\,\rm s$
cadence on which the BiSON data are stored.

The visibility of a mode in a frequency-power spectrum will depend
on the power and lifetime of the mode. We will now briefly discuss
how these properties affect the visibility.

\subsection{How the power and width of a mode affects its visibility}

The width of a mode, $\Delta$, is related to its lifetime, $\tau$,
by

\begin{equation}\label{equation[lifetime]}
    \Delta=\frac{1}{\pi\tau}.
\end{equation}

\noindent If the length of a timeseries is significantly longer than
the lifetime of a mode the timeseries will extend over several
realizations of the mode and so the modal peak will be resolved
across several bins in the frequency domain. However, if a mode's
lifetime is significantly longer than the length of a timeseries all
of the mode's power will be contained in a single bin\footnote{The
power can be split between 2 bins if the frequency of the mode is
not commensurate with the spectrum's frequency bins. Also the finite
length of a timeseries means that the mode will appear in a
frequency-power spectrum as a sinc squared function.}.

If the width of a mode can be resolved the power of that mode,
$V^2$, is given by

\begin{equation}\label{equation[power]}
    V^2=\frac{\pi}{2}T \Delta H,
\end{equation}

\noindent where $T$ is the length of the timeseries and $H$ is the
maximum power spectral density per bin. When the power spectrum is
fitted it is actually the maximum power spectral density, which
corresponds to the height, $H$, of the Lorentzian, that is
determined. The width, $\Delta$, and height, $H$, of the mode can
then be used to find the power of the mode. In this paper the
observed signal-to-noise ratio of a mode refers to the ratio of the
height of the most prominent spike across the width of the mode and
the mean level of the background around it. We define a spike as the
power (or amplitude) contained in one bin in a frequency-power (or
amplitude) spectrum. The height of the most prominent spike of a
resonant peak in a frequency-power spectrum will be greater than the
height of the fitted Lorentzian. This is because of the random
nature of the excitation of the modes, which means that the power
will have a $\chi^2$ 2 degrees of freedom distribution about the
underlying Lorentzian. Therefore, the power in some of the bins
across the width of the mode will be greater than the height of the
Lorentzian. These are the spikes that are most likely to be
detected.

Notice that it is the power, $V^2$, that we are going to extrapolate
and not the peak height of the mode. This is because the height,
$H$, of the mode is not a smooth function of frequency, as it
depends on whether the power of the mode is spread over more than
one bin or whether the mode's power is confined to one bin only. On
the other hand the power of a mode is a smooth function of frequency
and so can be extrapolated more easily.

A well-defined mode is one with a power and a width that are
sufficiently large for the mode to be detected easily, allowing the
shape of the mode in a frequency-power spectrum to be fitted
accurately. In the case of the Sun, when fitting a set of data that
spans $\sim8.5\,\rm yr$, well-defined modes have frequencies greater
than $\sim1500\,\rm\mu Hz$. Here we are concerned with modes at low
frequencies that cannot be observed clearly. Therefore it was
necessary to extrapolate the widths and powers of well-defined modes
to obtain estimates of the widths and powers of low-frequency modes.
We have, therefore, assumed that simple functional relationships,
which describe the variation of the parameters with frequency at
higher frequencies, persist to lower frequencies. In section
\ref{subsection[model predictions]} we have used predictions from a
stochastic excitation model to test the validity of this assumption.
The extrapolation can be performed in two different ways, each of
which will now be described in turn.

\subsection{Method 1: The ln-linear relationship}

\begin{figure}
  \centering
  \subfigure{\includegraphics[width=0.35\textwidth, clip]{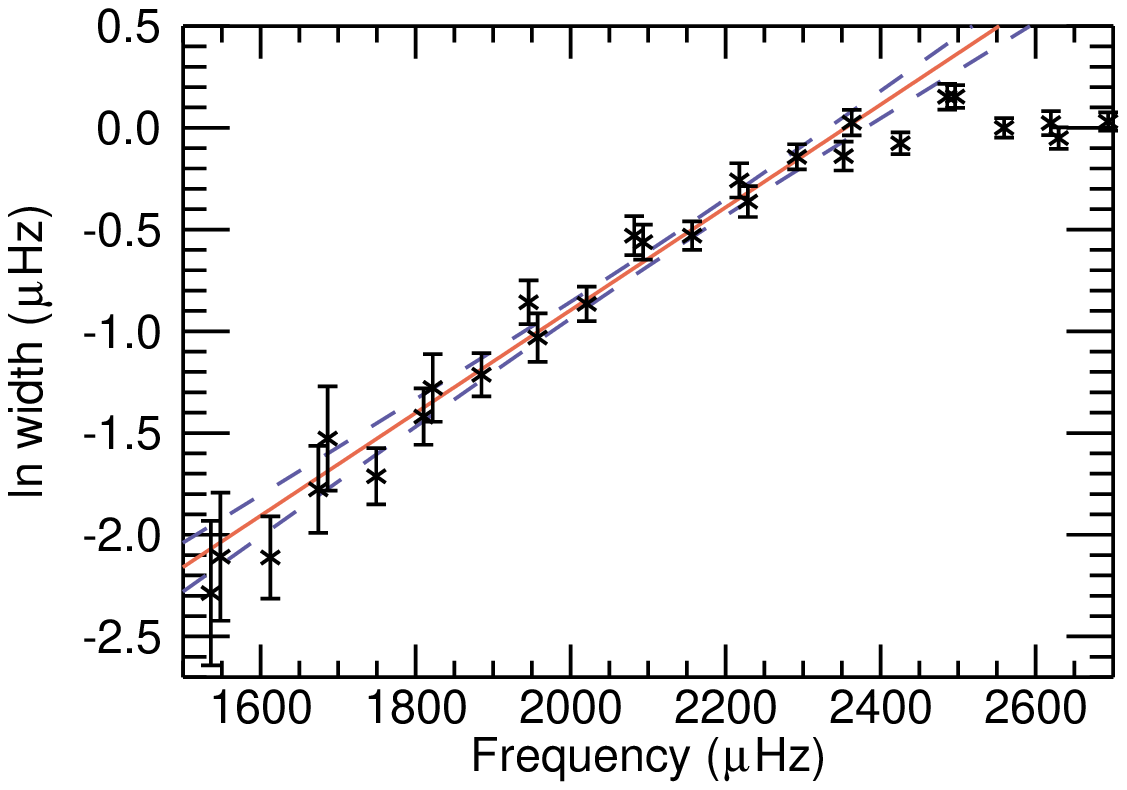}}\\
  \subfigure{\includegraphics[width=0.37\textwidth, clip]{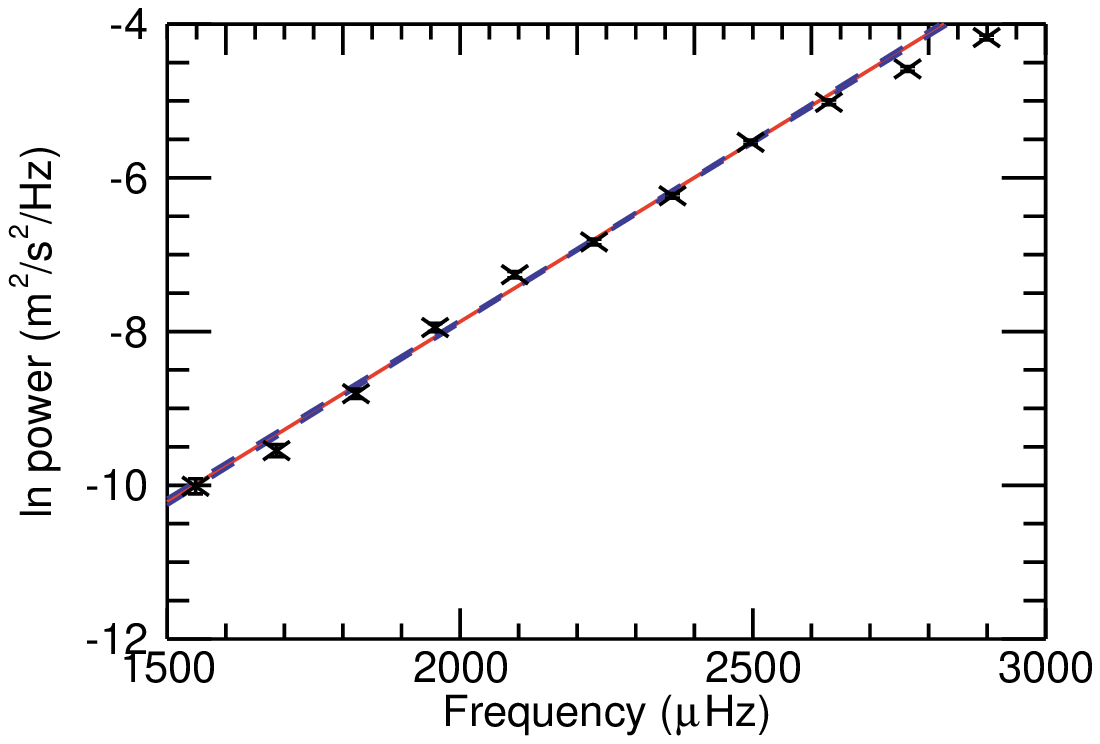}}\\
  \caption{Results for method 1. Top panel: The black crosses show how the natural logarithm
  of the observed width of a mode varies with frequency. The widths
  have been plotted for modes with $l=0$, 1 and 2. The errors of the widths are
  those associated with the fits. The red solid line shows the linear line
  of best fit for the widths at frequencies between
  $\sim1530\,\rm\mu Hz$ and $\sim2400\,\rm\mu Hz$. The blue dashed
  lines indicate the errors on this linear fit. Bottom panel: The black
  crosses show how the natural logarithm of a mode's power varies with
  frequency. The powers have been plotted for $l=0$ modes only.
  The red solid line shows the linear line
  of best fit for the powers at frequencies between
  $\sim1540\,\rm\mu Hz$ and $\sim2500\,\rm\mu Hz$. The blue dashed
  lines indicate the errors on this fit.}\label{figure[lgln]}
\end{figure}

For well-defined modes with frequencies below $\sim2400\,\rm\mu Hz$
an approximately linear relationship is observed between the natural
logarithm of a mode's width and its frequency (see the top panel of
Figure \ref{figure[lgln]}). The well-defined plateau in the mode
widths above $\sim2400\,\rm\mu Hz$ (see for example
\cite{Chaplin2005}) means that the linearity does not extend to
higher frequencies. The observed widths have been fitted using data
from modes with different degrees. In Sun-as-a-star observations
only low-$l$ modes can be clearly observed. Here widths from $l=0$,
1 and 2 modes have been used. The width of a mode is not independent
of degree as high-$l$ modes have a lower inertia than low-$l$ modes
and so high-$l$ modes have faster damping rates. However, the
difference in damping rates is minimal over the confined range of
$l$ used here.

The relationship between the natural logarithm of a mo\-de's power
and its frequency is also approximately linear for well-defined
modes with frequencies below $\sim2500\,\rm\mu Hz$ (see the bottom
panel of Figure \ref{figure[lgln]}). The Doppler velocity variations
that a mode exhibits over the solar disc are characterized by the
spherical harmonics that describe the mode, and so are reliant on
degree. The observed power is then dependent on the sensitivity of
the observing instrument to the distribution of the line-of-sight
velocities over the portion of the solar disc that is being
observed. The bottom panel of Figure \ref{figure[lgln]} shows the
results for $l=0$ modes only. The power is calculated from equation
\ref{equation[power]} using the width and the height of the best fit
Lorentzian for the mode. The width and height of a mode have a
negative correlation of $\sim0.95$ and so the resulting error bars
on the calculated powers are small.

For both the power and the width a linear, least-squares fit was
performed to determine the gradients and the zero-frequency
intercepts. The gradients and zero-frequency intercepts of these
fits and the errors associated with them can be seen in Table
\ref{table[m and c]}. This method of fitting the data will be
referred to as method 1.

\subsection{Method 2: The ln-ln relationship}
\begin{figure}
  \centering
  \subfigure{\includegraphics[width=0.35\textwidth, clip]{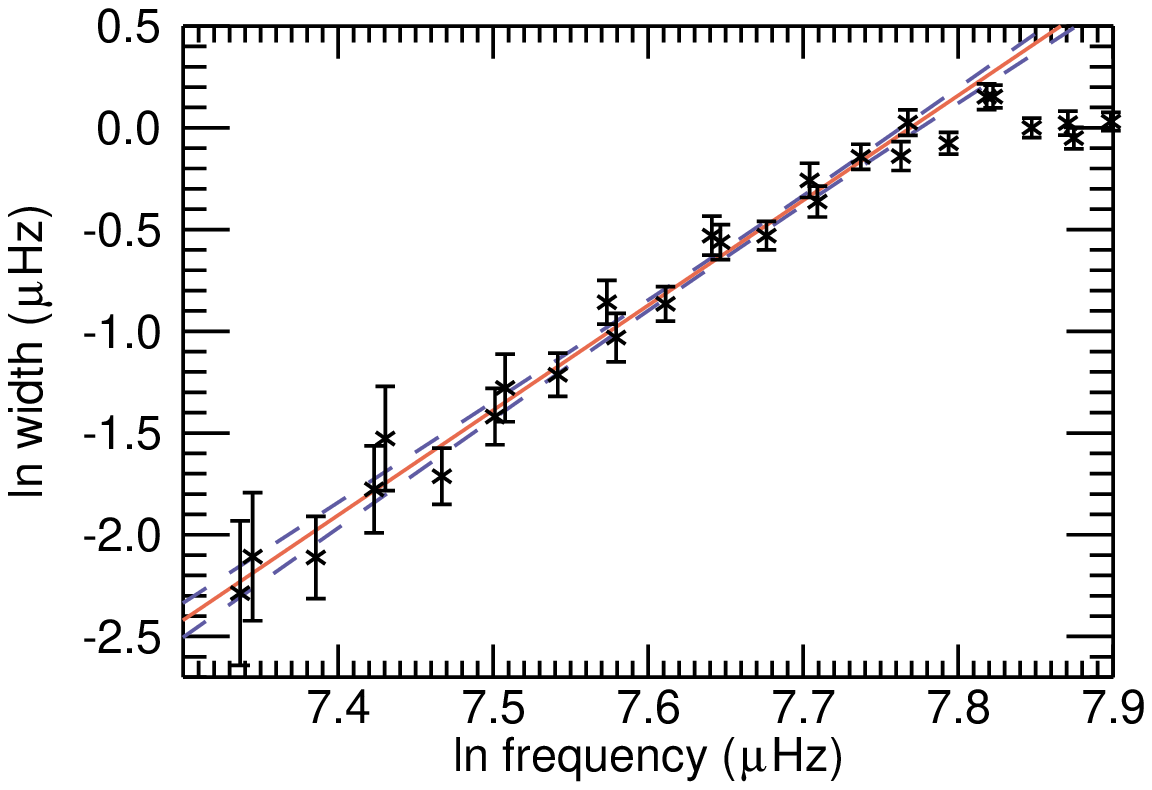}}\\
  \subfigure{\includegraphics[width=0.35\textwidth, clip]{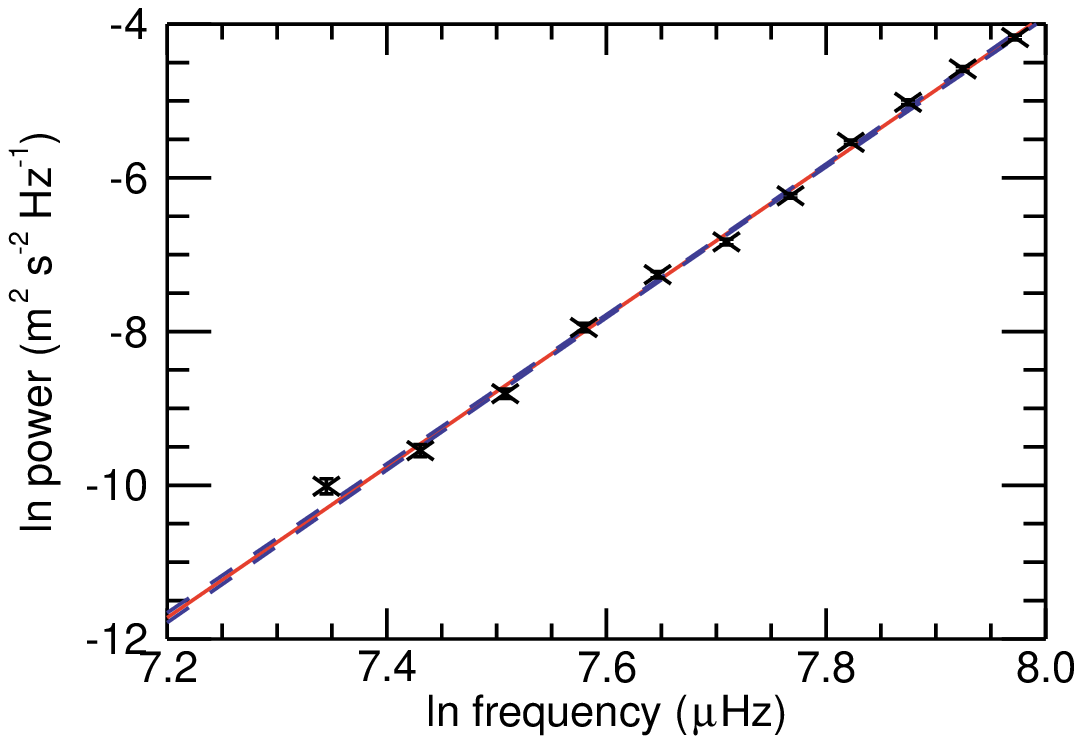}}\\
  \caption{Results for method 2. Top panel: The black crosses show how the natural logarithm
  of the observed width of a mode varies with the natural logarithm of its
  frequency. The errors bars shown for the widths are
  those associated with the fits. The widths
  have been plotted for modes with $l=0$, 1 and 2. Bottom panel: The black
  crosses show how the natural logarithm of the mode power varies
  with the natural logarithm of
  mode frequency. The powers have been plotted for $l=0$ modes only.
  As in Figure \ref{figure[lgln]} the red solid
  lines in both panels represent the linear best fits and the blue
  dashed lines represent the errors on the fits.}\label{figure[lglg]}
\end{figure}

An approximately linear relationship is also observed between the
natural logarithm of the width of a mode and the natural logarithm
of its frequency (see the top panel of Figure \ref{figure[lglg]}).
Once again the linearity is only found at frequencies below
$\sim2400\,\rm\mu Hz$. Likewise an approximately linear relationship
is found when the natural logarithm of the power of a mode is
plotted against the natural logarithm of the frequency of that mode
(see the bottom panel of Figure \ref{figure[lglg]}). As with method
1 a linear least squares fit was performed to find the gradients and
the zero-frequency intercepts for both the width and the power. The
gradients and zero-frequency intercepts resulting from fitting the
powers and widths in this manner can be seen in Table \ref{table[m
and c]}. This approach to fitting the data will be referred to as
method 2.

\begin{table*}
\centering \caption{Gradients and zero-frequency intercepts found by
least squares fits for method 1 and method 2. The reduced
$\chi_\nu^2$ value for each fit is also given.}\label{table[m and
c]}
\begin{center}
\begin{tabular}{cccc}
  \hline
  & Gradient & Zero-frequency Intercept & $\chi_\nu^2$ \\
  \hline
  Width &  & \\
  \hline
  Method 1 & $(2.5\pm0.1)\times10^{-3}$ & $-6.0\pm0.2$ & 0.91\\
  Method 2 & $(5.2\pm0.2)\times10^{0}$ & $-40\pm2$ & 0.70\\
  \hline
  Power & & \\
  \hline
  Method 1 & $(4.68\pm0.06)\times10^{-3}$ & $-17.2\pm0.1$ & 6.5\\
  Method 2 & $(9.8\pm0.1)\times10^{0}$ & $-82.3\pm0.9$ & 6.9\\
  \hline
\end{tabular}
\end{center}
\end{table*}


\subsection{Extrapolating method 1 and method 2 to low frequencies}
To determine the value of the width and power of a mode at very low
frequencies the gradients and zero-frequency intercepts were used to
extrapolate the linear relationships from method 1 and method 2 to
lower frequencies. The widths found using both method 1 and method 2
for $l=0$ modes, in the frequency range $800$ to $1450\,\rm\mu Hz$,
can be seen in Table \ref{table[fitted widths]}; and the powers for
the same modes, also found using both method 1 and method 2, can be
seen in Table \ref{table[extrapolated powers]}. Some of these modes
have not yet been observed and so we have used the mode frequencies
predicted by the Saclay Seismic Model \cite{Turck-Chieze2001}.


\begin{table}
\centering \caption{Predicted widths of various $l=0$ modes found by
extrapolating both the relationship found in method 1 and the
relationship found in method 2.}\label{table[fitted widths]}
\begin{center}
\begin{tabular}{ccc}
  \hline
  \small{Frequency} & \small{Width Found Using} & \small{Width Found Using} \\
  \small{($\rm\mu Hz$)} & \small{Method 1 ($\rm\mu Hz$)} & \small{Method 2 ($\rm\mu Hz$)} \\
  \hline
  1407.627 & $0.091\pm0.008$ & $0.07\pm0.01$\\
  1263.524 & $0.064\pm0.006$ & $0.039\pm0.008$\\
  1118.15 & $0.044\pm0.005$ & $0.021\pm0.005$\\
  972.745 & $0.031\pm0.004$ & $0.010\pm0.003$\\
  825.365 & $ 0.021\pm0.003$ & $0.0044\pm0.002$\\
  \hline
\end{tabular}
\end{center}
\end{table}



\begin{table*}
\centering \caption{Predicted powers of various $l=0$ modes found by
extrapolating the relationship found in method 1 and the
relationship found in method 2.}\label{table[extrapolated powers]}
\begin{center}
\begin{tabular}{ccc}
  \hline
  \small{Frequency}  & \small{Power Found Using}  & \small{Power Found Using} \\
  \small{($\rm\mu Hz$)} & \small{Method 1 ($\rm m s^{-2}$)} & \small{Method 2 ($\rm m s^{-2}$)} \\
  \hline
  1407.627 & $(2.38\pm0.09)\times10^{-5}$ & $(1.32\pm0.08)\times10^{-5}$\\
  1263.524 & $(1.21\pm0.05)\times10^{-5}$ & $(0.46\pm0.03)\times10^{-5}$\\
  1118.15 & $(0.61\pm0.03)\times10^{-5}$ & $(0.14\pm0.01)\times10^{-5}$\\
  972.745 & $(0.31\pm0.02)\times10^{-5}$ & $(0.035\pm0.002)\times10^{-5}$\\
  825.365 & $(0.16\pm0.01)\times10^{-5}$ & $(0.0078\pm0.0008)\times10^{-5}$\\
  \hline
\end{tabular}
\end{center}
\end{table*}


As can be seen the widths extrapolated using method 2 are
consistently smaller than the widths extrapolated using method 1.
The difference between the estimated width of a particular mode
increases as the frequency of the mode decreases. In fact the widths
inferred by the two methods do not agree to within their associated
error bars at any frequency. Below $1000\,\rm\mu Hz$ the widths
predicted by method 1 are $\sim3$ times larger than the widths
predicted by method 2. Additionally method 2 consistently predicts
lower powers than method 1. Again the difference between the two
extrapolations increases as the frequency of the mode decreases and
at no time do the powers predicted by the two methods agree to
within their respective error bars. At the higher frequencies the
powers predicted by method 2 are just under half the powers
predicted by the method 1. However, below $1000\,\rm\mu Hz$ the
powers predicted by method 1 are approximately a factor of 10 larger
than the powers predicted by method 2. Clearly there is a
significant discrepancy between the powers and widths predicted by
the two methods.

Table \ref{table[m and c]} gives the reduced $\chi_\nu^2$ values of
the linear fits. The lower the reduced $\chi_\nu^2$ value the better
the fit is at representing the data. However, if the reduced
$\chi_\nu^2$ value is significantly less than unity either the model
used to fit the data is too complicated or the errors on the data
have been overestimated. The $\chi_\nu^2$ values are very similar
for each method with method 2 providing a slightly better fit for
the widths but method 1 giving a better representation of the
powers. Both methods appear to give very good fits to the width. The
values of the reduced $\chi_\nu^2$ determined for the linear power
fits are large because of the small error bars associated with the
observed powers.

Monte Carlo simulations were performed using the results of both
extrapolations. These simulations will be described in Section
\ref{section[simulations]}. However, before we describe the
simulations it is interesting to compare the extrapolated parameters
with theoretical predictions that can be found using a stochastic
excitation model.

\subsection{Comparing the results of the extrapolation with
solar excitation model predictions}\label{subsection[model
predictions]}

\begin{figure}
  \centering
  \includegraphics[width=0.35\textwidth, clip]{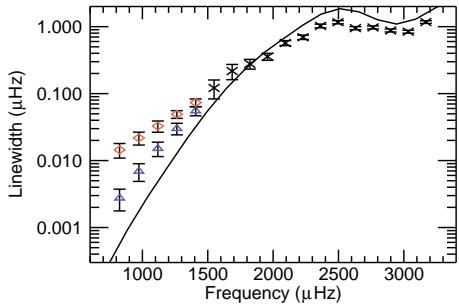}\\
  \caption{A comparison between observed, modelled and
  extrapolated linewidths. The solid line represents the modelled
  linewidths, the black crosses represent the linewidths found by
  fitting BiSON data, the red diamonds show the widths found using
  extrapolation method 1 and the blue triangles show the widths
  found using extrapolation method 2.}\label{figure[model widths]}
\end{figure}

Theoretical model mode damping and excitation rates can be
calculated using various analytical models, which describe the
interaction of the convection and the oscillations. The power of a
mode, $V^2$, is given by
\begin{equation}\label{equation[model power]}
    V^2=\frac{P}{2\pi\Delta I}
\end{equation}
\noindent where $P$ is the energy supply rate, $I$ is the mode
inertia and $\Delta$ is, as defined previously, the width of the
mode in a frequency-power spectrum. In what follows we will compare
the powers and linewidths that have been predicted by one such
stochastic excitation model with the linewidths and powers that are
observed in and can be extrapolated from BiSON data. Chaplin et al.
(2005)\nocite{Chaplin2005} calculated model energy supply rates and
linewidths using the Cambridge stochastic excitation and damping
codes. Figure \ref{figure[model widths]} shows that the model widths
drop off more rapidly than both the observed and extrapolated widths
at low frequencies. This is a known, and as yet unresolved problem,
with the modelling of the linewidths (see for example
\cite{Chaplin2005}).

\begin{figure}
  \centering
  \subfigure{\includegraphics[width=0.35\textwidth, clip]{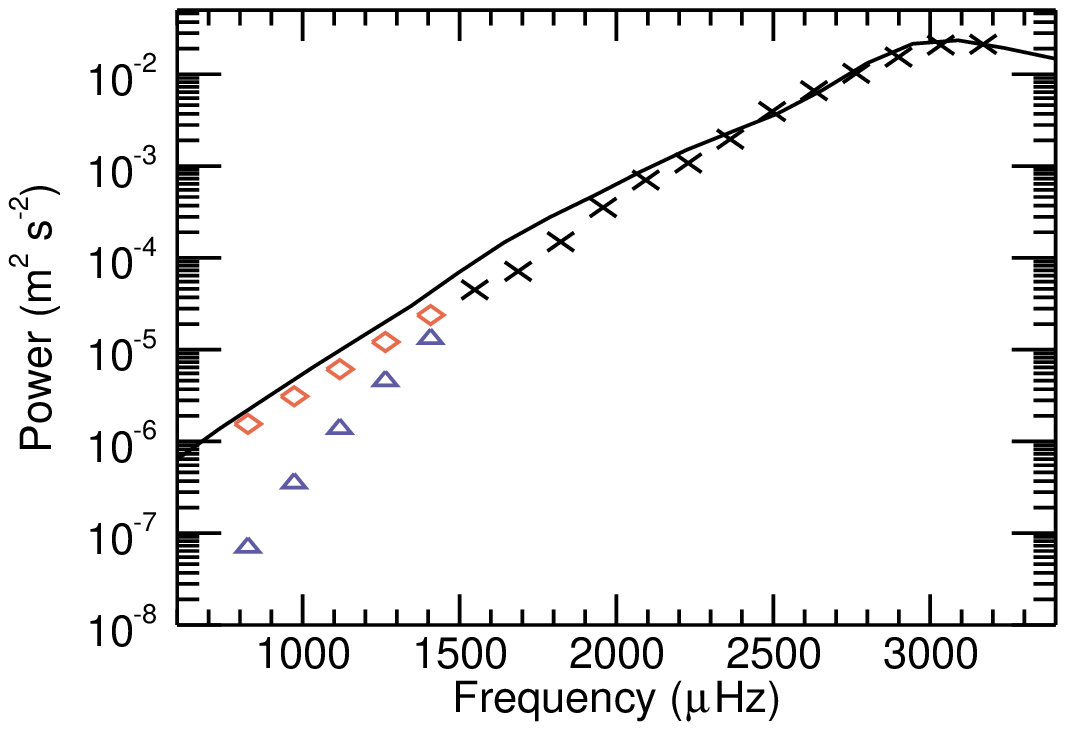}}\\
  \subfigure{\includegraphics[width=0.35\textwidth, clip]{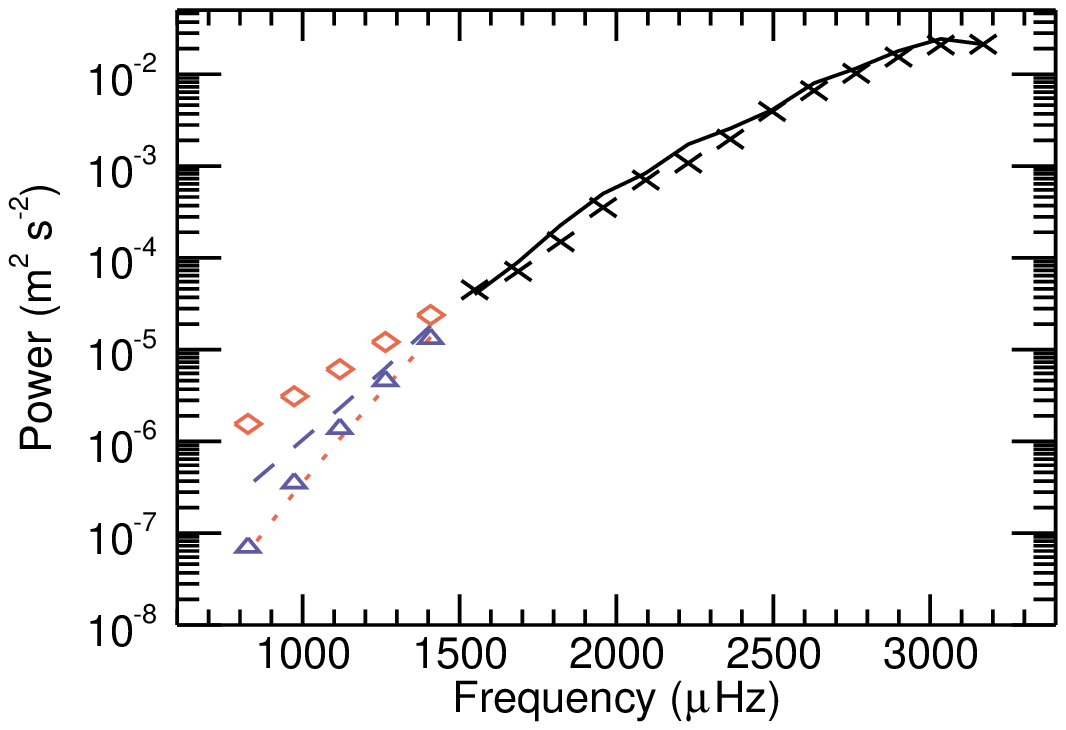}}\\
  \caption{Top panel: A comparison between the observed, modelled and
  extrapolated powers. The solid line represents the modelled
  powers, the black crosses represent the powers found by
  fitting BiSON data, the red diamonds show the powers found using
  extrapolation method 1 and the blue triangles show the powers
  found using extrapolation method 2. Bottom panel: A comparison
  between the observed, modelled and extrapolated powers, however the
  model powers have been found using the observed widths. The
  symbols and lines have the same definitions as in the top panel.
  Also plotted are the model powers found using the widths
  estimated by extrapolation method 1 and the model powers
  calculated using the widths extrapolated using method 2.}
  \label{figure[model powers]}
\end{figure}

The top panel of Figure \ref{figure[model powers]} shows a
comparison between the model, observed and extrapolated powers. As
we are interested in the frequency dependence of the power the model
powers have been scaled to ensure that the maximum model power and
the maximum observed power are equal. The model powers decrease less
rapidly than the powers observed in the BiSON data. Therefore the
powers estimated by both method 1 and method 2 are smaller than the
model powers at low frequencies. However, this is understandable as
the model widths decrease more rapidly than the observed widths.
Chaplin et al. (2005)\nocite{Chaplin2005} note that the model widths
are too narrow at low frequencies, which implies that the
extrapolations performed tend in the right direction.

It is also of interest to find the model powers using the observed
widths. The bottom panel Figure \ref{figure[model powers]} shows
that using the observed widths produces a far better agreement with
the powers observed in the BiSON data. This indicates that the
majority of the discrepancy between the observed and model powers
seen in the top panel of Figure \ref{figure[model powers]} is due to
the smaller model widths. The extrapolated powers also appear to
agree better with the model values calculated using the observed
widths (bottom panel of Figure \ref{figure[model powers]}). However,
it is difficult to tell which extrapolation method produces the best
agreement with the model powers. Also plotted in the bottom panel of
Figure \ref{figure[model powers]} are the model powers that are
found when the widths estimated by both extrapolation methods are
used. The agreement between the extrapolated powers and the model
powers is poor when method 1 is used. However, the agreement between
the extrapolated and model powers is better when method 2 is used.

\section{Simulating the modes}\label{section[simulations]}
In this section we describe how we made artificial p-mode
timeseries. These timeseries were analyzed, and results on
detections of their artificial modes compared with the results of
Broomhall et al. (2007), on real p-mode data. In Broomhall et
al.\nocite{Broomhall2007} BiSON and GOLF frequency-amplitude spectra
were searched for coincident prominent structures that occurred at
the same frequency in each spectrum. Here we needed to simulate
pairs of timeseries, which both contained the signal from a mode and
normally distributed noise. The statistical tests described in
Broomhall et al. (2007)\nocite{Broomhall2007} were then applied to
the simulated data to determine how often the simulated modes could
be detected.

Modes of a given frequency and lifetime (width) were simulated by
randomly exciting an oscillator that was damp\-ed over the correct
timescale. The damping time was predicted using equation
\ref{equation[lifetime]} and the extrapolations described in Section
\ref{section[extrapolations]} for linewidth. The simulations
produced timeseries that contained the signal from a simulated mode.
The total power of the simulated mode was scaled to the power
predicted by the simulations.

Two timeseries containing normally distributed random noise were
created. The signal from two sets of contemporaneous real
Sun-as-a-star data taken by different instruments, such as BiSON and
GOLF, contains some coherent noise. This noise is solar in origin
and comes from the solar granulation. The level of coherent noise is
frequency dependent and at around $1000\,\rm\mu Hz$ the coherency
between BiSON and GOLF data is $\sim0.1$. Therefore 10\% of the
noise in the two simulated timeseries was set so it was common to
both sets of simulated data. The timeseries containing the signal
from the simulated mode was then added to each noise timeseries.

Solar modes are excited stochastically by turbulence in the
convection zone, which is caused by the solar granulation. As the
coherent noise in two sets of Sun-as-a-star data is from the
granulation it is possible that the coherent noise found between the
BiSON and GOLF data is also correlated to the amplitude of
excitation of the mode. In the simulations the amplitude of the
excitation was determined by a normally distributed array. For some
of the simulations the correlated noise that was added to the
simulated data was taken to be the array that determined the
amplitude of the excitation of the mode. Simulations were also
performed when the noise that was coherent between the two simulated
sets of data was independent of the amplitude of excitation of the
mode.

The simulations have been performed for the $l=0$ modes between
$\sim800$ and $\sim1450\,\rm\mu Hz$, the extrapolated properties of
which are given in Tables \ref{table[fitted widths]} and
\ref{table[extrapolated powers]}. The timeseries were simulated to
contain 2,211,120 points with a cadence of 120s to be consistent
with the sets of BiSON and GOLF data used in Broomhall et al.
(2007)\nocite{Broomhall2007}. The simulated and BiSON timeseries
cover the same length in time ($\sim8.5\,\rm yrs$), despite having
different cadences and containing a different number of points. This
means that the power of a mode in the simulated and observed
timeseries is the same. Furthermore, the number of frequency bins
that cover the width of a mode in the simulated frequency-power
spectra is the same as in the fitted BiSON spectrum. Therefore, the
height of the mode in a simulated frequency-power spectrum will be
the same as the height predicted by the extrapolations. The level of
noise in the simulated spectra was scaled to mimic the mean level of
noise observed in the BiSON frequency-power spectrum in the
$100\,\rm\mu Hz$ surrounding the frequency of the mode that was
simulated.

It should be noted that we are considering low-frequency modes,
which are relatively insensitive to the effects of the solar
activity cycle. We would therefore not expect solar cycle effects to
have a significant impact on results, and so solar cycle effects
were neglected in the simulations.

Each extrapolated power and width has an error associated with it.
To obtain an upper limit on the fraction of times a simulated mode
could be detected Monte Carlo simulations were performed using the
$\pm1\sigma$ error values of the extrapolated widths and powers.

Once timeseries containing a mode and noise had been created several
tests were performed on the resulting frequen\-cy-amplitude spectra
to determine whether the mode could be detected or not. For each
mode 1000 pairs of timeseries were simulated. Each pair of simulated
timeseries corresponds to a different realization. The number of
pairs of spectra in which a mode was detected was then counted.
Various statistical tests were used to detect the modes. We will now
outline each of these tests in turn.

\section{Statistical Tests}\label{section[stat tests]}
The tests performed are based on the statistics described in detail
in Chaplin et al. (2002)\nocite{Chaplin2002} and Broomhall et al.
(2007)\nocite{Broomhall2007}. Here we summarize the statistical
tests that were used to search the simulated spectra. The simplest
test involves sear\-ching for a single prominent spike that is above
a given threshold level in the same bin in each of the two simulated
frequency-amplitude spectra. The threshold level for detection was
set at a 1 per cent chance of getting at least one false detection
anywhere in $100\,\rm\mu$Hz. This level and frequency range were
chosen to maintain consistency with the detection methods employed
in Broomhall et al. (2007)\nocite{Broomhall2007}. The probability of
observing such a `pair of spikes' takes proper account of the level
of common noise present in the two frequency-amplitude spectra as
this affects the probability that any detection is due to noise. A
detection was considered to have been made if it was positioned
within one linewidth of the input frequency. The linewidths were
determined by the extrapolated values and so varied between modes.
As the modes are damped it is possible to take advantage of the
width the mode may exhibit in frequency. Furthermore, if the
frequency of a mode is not commensurate with the frequency bins of
the spectrum the power of the mode may be spread over more than one
bin. The second test involves searching for two prominent spikes in
the same consecutive frequency bins of each frequency-amplitude
sp\-ectrum. Both spikes must lie within one peak width of the mode's
input frequency for a detection to be counted. The third test
performed also searched for two prominent spikes but this time the
spikes did not need to be in consecutive bins. These two spikes are
known as a two-spike cluster. The bins occupied by the two spikes
could lie up to twice the predicted peak width apart. However each
spike had to lie less than one linewidth from the mode's input
frequency for a detection to be counted. The two prominent spikes in
the cluster needed to be in the same bins in each
frequency-amplitude spectrum. This test was then extended to search
for clusters containing three, four and five prominent spikes. The
highest and lowest frequency spikes in the cluster had to be
separated by less than twice the width of the mode. Furthermore,
each prominent spike in the cluster had to be in the same frequency
bin in each frequency-amplitude spectrum. All spikes in the clusters
needed to lie within one linewidth of the mode's input frequency for
a detection to be counted.

The simulated spectra were searched to determine how often these
statistical tests were passed for each mode. The results of these
tests will now be described.

\section{Results of the Simulations}\label{section[results]}

\begin{figure}
  \centering
  \includegraphics[width=0.35\textwidth, clip]{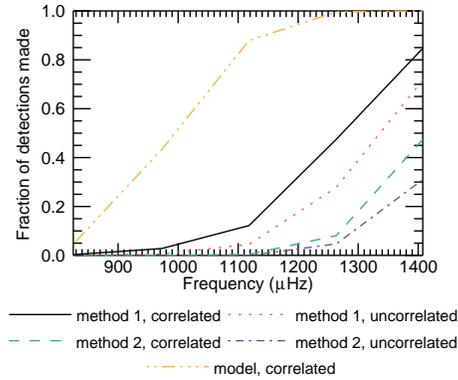}\\
  \caption{Comparison between the fraction of modes detected when different input parameters were used.}
  \label{figure[method comp]}
\end{figure}

Figure \ref{figure[method comp]} shows the fraction of the 1000
simulations that were performed in which the mode being simulated
was detected when looking for a single spike in the same bin in each
spectrum. The results show that using the parameters predicted by
method 1 leads to more detections than the parameters estimated by
method 2. This is expected as the powers predicted by method 1 are
significantly larger than the powers predicted by method 2. More
detections are made when the noise that is coherent between the two
sets of data is also correlated to the amplitude of the excitation
of the mode. This is also understandable. If, for example, at a
point in time the level of solar noise is larger than average the
amplitude of the excitation of the mode will also be larger than
average enabling it to potentially remain distinguishable from the
noise. However, all of the simulations that use the extrapolated
values predict that the number of detections at low frequencies will
be small. Also plotted on Figure \ref{figure[method comp]} are the
results of simulations performed using the model widths and powers
predicted by the stochastic excitation model described in Chaplin et
al. (2005)\nocite{Chaplin2005}. The model powers have been found
using the model widths rather than the observed widths. Clearly more
detections are made when the model parameters are used.

\begin{figure*}
  \centering
  \subfigure[Method 1]
  {\includegraphics[width=0.35\textwidth, clip]{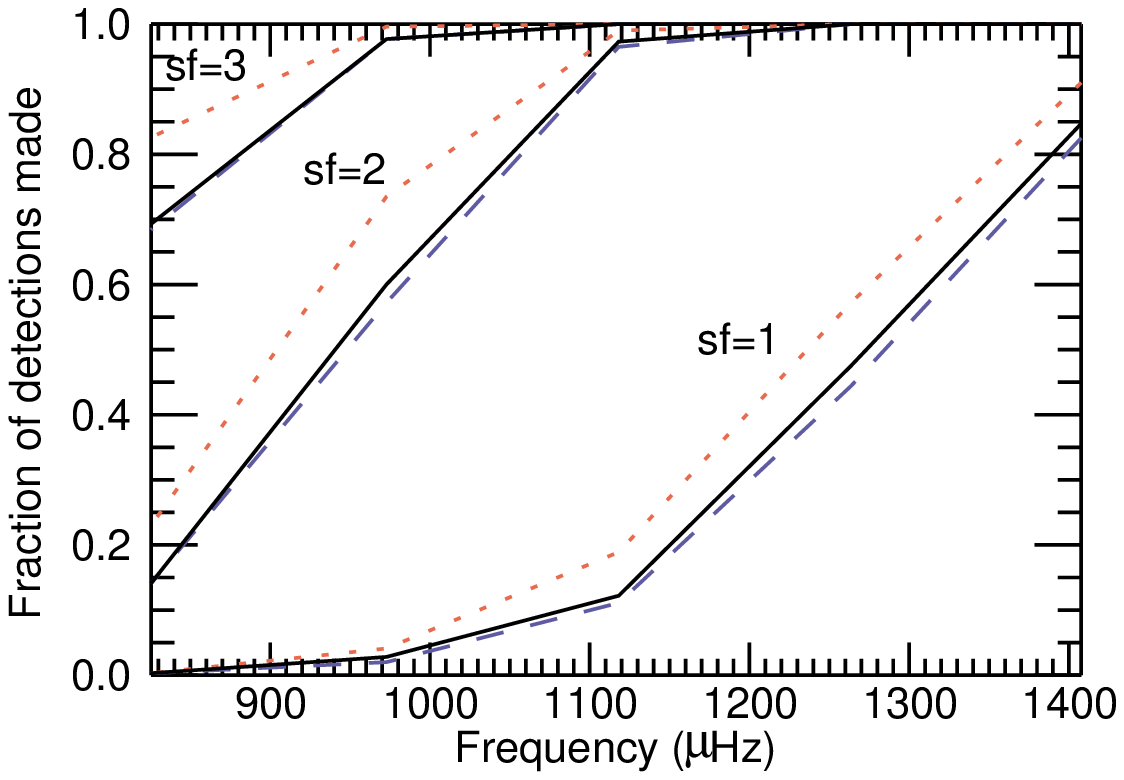}}
  \hspace{1cm}
  \subfigure[Method 2]
  {\includegraphics[width=0.35\textwidth, clip]{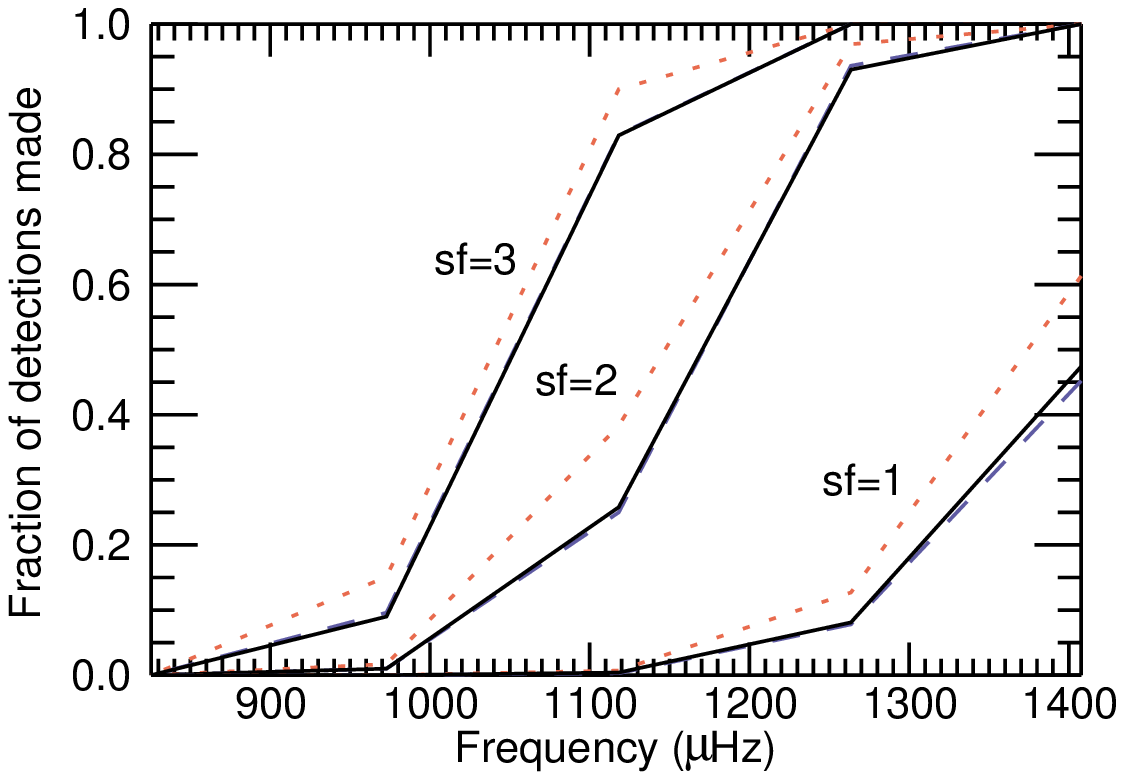}}
  \caption{Results of searching for a single spike when the noise is correlated with the excitation amplitude. The black solid
  lines show the results for the actual extrapolated values. The red dashed
  lines show the results when the maximum power and minimum widths
  are used in the simulations. The blue dashed line shows the results of the
  simulations that used the maximum powers and the maximum widths.
  The three clusters of lines are present as for each set of
  parameters the power was scaled by a different amount. The
  clusters labelled sf=1 represent the true predicted values. The
  clusters labelled sf=2 show the results when the power
  was increased by a factor of 2 and
  the clusters labelled sf=3 are the results when the power was
  increased by a factor of 3.
  }\label{figure[single_results]}
\end{figure*}

Figure \ref{figure[single_results]} plots results from the spike
tests where the coherent noise was correlated to the mode
excitation. Each panel of Figure \ref{figure[single_results]} shows
three clusters of lines. The solid, black lines in the clusters
labelled sf=1 represent the results from simulations where the power
and width of the input mode were predicted by the respective
extrapolation methods. The results show that the number of
detections decreases rapidly as we move to lower frequencies. As
each of the statistical tests are looking for prominent spikes it is
the height the mode exhibits in the frequency-amplitude/power
spectra that determines whether or not it can be detected. The
height of a mode in a frequency-power spectrum is proportional to
$V^2/\Delta$ (see equation \ref{equation[power]}). The upper limit
on the number of detections is given by the case when the maximum
power and minimum width is used. Using the maximum power and minimum
width means that the height of the mode, $H$, is, potentially,
larger and so the signal-to-noise ratio of the mode should be
increased. It is therefore understandable that using the maximum
power and minimum width leads to more detections. When the
simulations were performed using the maximum width the power is
spread over a larger number of bins and so this negates the effect
on the signal-to-noise ratio of the mode of increasing its power.

Simulations were also performed to investigate the effect of
increasing the observed signal-to-noise ratio. The power given to
each simulated mode was increased by factors of 2 (labelled sf=2 in
each panel of Figure \ref{figure[single_results]}) and 3 (labelled
sf=3 in each panel of Figure \ref{figure[single_results]}). This
significantly increases the number of detections made at low
frequencies. The fraction of spectra in which the $l=0$, $n=6$ mode
was detected when the scale factor was 1 was at most 0.05. However,
the fraction of detections made when the scale factor was 2 was 0.60
and when the scale factor was 3 the fraction of simulated spectra in
which the $l=0$, $n=6$ mode was detected was 0.98.

\section{Discussion}\label{section[discussion]}
Clearly the visibility of the modes drops off at low frequencies.
The most optimistic results are achieved when the extrapolation is
performed using method 1 and the common noise is correlated to the
amplitude of the excitation of the mode. Clearly more detections are
made when the stochastic excitation model parameters are used.
However, these results can only be treated as an upper limit as we
have already seen that the model underestimates the widths of the
modes. This not only leads to the power of the modes being
overestimated but also means that the height of a mode is increased,
and so a mode is more likely to be detected.

For both extrapolation methods, irrespective of whether the noise is
correlated or uncorrelated with the excitation amplitude, the
fraction of detections at $\sim1000\,\rm\mu Hz$ is less than 0.1. On
the assumption that method 1 produces a more accurate prediction of
the mode's powers and widths than method 2 the simulations imply
that if the power signal-to-noise ratio can be increased by a factor
of 2 a significantly higher proportion of modes could be detected.
It would therefore be beneficial to continue efforts to try and
improve the quality of the data further.

The total background continuum is a combination of instrumental
noise, solar noise and, for Earth-based instruments such as the
BiSON network, a small amount of atmospheric noise. If the
atmospheric and instrumental noise can be reduced this would
increase the signal-to-noise ratio of a mode, making it easier to
detect. The limiting factor to the improvement that can be made to
the data quality is the amount of solar noise present. However, the
mean power of the BiSON data is always more than twice the power of
solar noise predicted by the Harvey model and the characteristic
parameters for the granulation found by \cite{Elsworth1994}.
Therefore it is theoretically possible to significantly reduce the
level of noise in this frequency range.

It is possible to compare the results of these simulations to the
observed candidates found in Broomhall et al.
(2007)\nocite{Broomhall2007}. According to the simulations the
$l=0$, $n=9$ mode at $\sim1407\,\rm\mu Hz$ should be detected in the
vast majority of spectra and this mode is observed to be very
prominent in the BiSON and GOLF data used in Broomhall at
al.\nocite{Broomhall2007}. The $l=0$, $n=8$ mode at
$\sim1263\,\rm\mu Hz$ is also detected by Broomhall et
al.\nocite{Broomhall2007}. The simulations indicate that the
fraction of time that this mode should be detected is at least 0.6.
The $l=0$, $n=7$ mode at $\sim1118\,\rm\mu Hz$ and the $l=0$, $n=5$
mode at $\sim825\,\rm\mu Hz$ are not detected in the BiSON and GOLF
data used in Broomhall et al.\nocite{Broomhall2007}. The fraction of
simulated spectra in which the $l=0$, $n=7$ mode was detected was,
at most, 0.35, while the fraction of simulations in which the $l=0$,
$n=5$ mode was detected was less than 0.06. Therefore it is
reasonable to expect that the modes might not be detected in the
BiSON and GOLF data.

It should of course be noted here that all of these results are
based on the assumption that the simple relationships for the mode
powers and widths observed in well-defined modes are still valid at
low frequencies. The solar excitation model suggests that this may
not be an unreasonable assumption, although it is difficult to
determine which extrapolation method predicts the lifetimes and
powers more accurately.

The $l=0$, $n=6$ mode is significantly more prominent in the
observations than the simulations. Therefore this mode will now be
considered in more detail.

\subsection{The $l=0$, $n=6$ mode}
The $l=0$, $n=6$ mode is observed to be prominent in Sun-as-a-star
data. However, some modes with higher frequencies remain undetected.
The most optimistic simulations imply that the fraction of spectra
in which the $l=0$, $n=6$ mode should be detected is less than 0.05.
In the simulations performed here, for method 1, with noise
correlated to the amplitude of the excitation, the average simulated
signal-to-noise ratio in amplitude was $\sim2.1$. However, the
observed amplitude signal-to-noise ratio in the GOLF and BiSON data
examined in Broomhall et al. (2007) was greater than $3.4$. In fact,
the fraction of simulations in which the signal-to-noise ratio of
the mode was greater than 3.4 was 0.014. Therefore the simulations
poorly represent the observed properties of this mode. It should be
noted that the amplitude signal-to-noise ratio of other modes that
were detected in Broomhall et al. are well represented by the
simulations in this paper. For example, the $l=0$, $n=9$ mode was
observed to have a signal-to-noise ratio of 3.1 while the average
signal-to-noise ratio produced by the simulations ranged from 2.9 to
3.7 depending on which extrapolation method was used and whether the
noise was correlated to the excitation amplitude.

When the mean amplitude signal-to-noise ratio of the simulated
$l=0$, $n=6$ mode was set as 3.4 the fraction of spectra in which
the mode was detected increased to $\sim0.6$. A signal-to-noise
ratio of 3.4 is $\sim3\sigma$ from the mean amplitude
signal-to-noise found in the original simulations where
extrapolation method 1 (ln-linear) is used and the noise is
correlated to the excitation amplitude. Therefore, the simulations
and the observations are not so different that new physics is
required to explain the observations. It is more likely that at some
point in time the mode has been randomly excited to a larger
amplitude because of the stochastic nature of mode excitation.

\section*{Acknowledgements}

This paper utilizes data collected by the Birmingham
Solar-Oscillations Network (BiSON). The calibrated BiSON data that
were used in this paper were prepared for the \emph{Phoebus}
collaboration. We would like to thank all of the members of
\emph{Phoebus} and the ISSI (http://www.issibern.ch/) for supporting
the \emph{Phoebus} collaboration. The authors would like to thank G.
Houdek for providing model excitation and damping rates. This work
was supported by the European Helio- and Asteroseismology Network
(HELAS), a major international collaboration funded by the European
Commission's Sixth Framework Programme.

BiSON is funded by the Science and Technology Facilities Council
(STFC). We thank the members of the BiSON team, colleagues at our
host institutes, and all others, past and present, who have been
associated with BiSON. The authors also acknowledge the financial
support of STFC.

\end{document}